\documentclass{ws-ijmpa}
\usepackage{amsmath}
\usepackage{amssymb}
\usepackage{graphicx}

\def\issue(#1,#2,#3){{\bf #1}, #2 (#3)} 

\def\opcit(#1){ {\em op. cit.}, #1}

\def\ARNPS(#1,#2,#3){Ann.\ Rev.\ Nucl.\ Part.\ Sci.\ \issue(#1,#2,#3)}
\def\CPC(#1,#2,#3){Comp.\ Phys.\ Comm.\ \issue(#1,#2,#3)}
\def\CIP(#1,#2,#3){Comput.\ Phys.\ \issue(#1,#2,#3)}
\def\EPJC(#1,#2,#3){Eur.\ Phys.\ J.\ C\ \issue(#1,#2,#3)}
\def\IEEETNS(#1,#2,#3){IEEE Trans.\ Nucl.\ Sci.\ \issue(#1,#2,#3)}
\def\NP(#1,#2,#3){Nucl.\ Phys.\ \issue(#1,#2,#3)}
\def\NIM(#1,#2,#3){ Nucl.\ Instrum.\ and Meth.\ \issue(#1,#2,#3)}
\def\PL(#1,#2,#3){Phys.\ Lett.\ \issue(#1,#2,#3)}
\def\PRD(#1,#2,#3){Phys.\ Rev.\ D \issue(#1,#2,#3)}
\def\PRL(#1,#2,#3){Phys.\ Rev.\ Lett.\ \issue(#1,#2,#3)}
\def\SJNP(#1,#2,#3){Sov.\ J. Nucl.\ Phys.\ \issue(#1,#2,#3)}
\def\ZPC(#1,#2,#3){Z.\ Phys.\ C \issue(#1,#2,#3)}
\def\PAN(#1,#2,#3){Phys.\ Atom.\ Nucl.\ \issue(#1,#2,#3)}
\begin{document}

\markboth{Kevin Stenson}
{Pentaquark searches at FOCUS}

%
\catchline{}{}{}{}{}
%

\title{Pentaquark searches at FOCUS}
\author{\footnotesize Kevin Stenson\footnote{on behalf of the FOCUS Collaboration (http://www-focus.fnal.gov/)}}
\address{Department of Physics, University of Colorado, Campus Box 390\\
Boulder, CO 80309, USA}

\maketitle

\begin{abstract}
We find no evidence for high-energy photoproduction of pentaquarks at 1540\,MeV$\!/c^2$, 1862\,MeV$\!/c^2$, 
or 3099\,MeV$\!/c^2$ using decay modes $pK_S^0$, $\Xi^-\pi^-$, and $D^{(*)-}p$, respectively.
\end{abstract}

\section{Introduction}

A 4--7\,$\sigma$ significant pentaquark with a mass of
$\sim$1540\,MeV$\!/c^2$ decaying to $pK_S^0$ or $nK^+$ 
has been reported by ten
experiments\,$^{\textrm{1--10}}$.
Combining the
mass measurements of these experiments we find $M = 1533.6 \pm 1.2$\,MeV$\!/c^2$. 
The $\chi^2/dof$ for this averaging is $38.2/9$ giving a confidence level of $1.6 \times 10^{-5}$, a 
5.2\,$\sigma$ problem.  
Pentaquarks with two strange quarks\,\cite{na49} and with a charm quark\,\cite{h1} have also been reported.

FOCUS ran during the 1996--7 fixed-target run at Fermilab.  
A photon beam from brehmsstrahlung of a 300 GeV electron and positron beam
impacts BeO targets.  16 silicon strip planes provide vertexing and tracking.
Charged particles are tracked and momentum analyzed as they pass
through up to two dipole magnets and up to five sets of multiwire proportional chambers.  
Three \v{C}erenkov counters, two EM calorimeters, and two muon detectors identify particles.
A hadronic trigger requiring $\sim$25\,GeV of energy passed 7 billion events for reconstruction.
Thus, these events are well above threshold for pentaquark production.  Charge
conjugates are assumed for these analyses and all pentaquarks are assumed to decay strongly.

\section{Search for $\Theta(1540)^+\!\rightarrow\! p K_S^0$}
\begin{figure}
\centerline{\includegraphics[width=5.0in,height=1.61in]{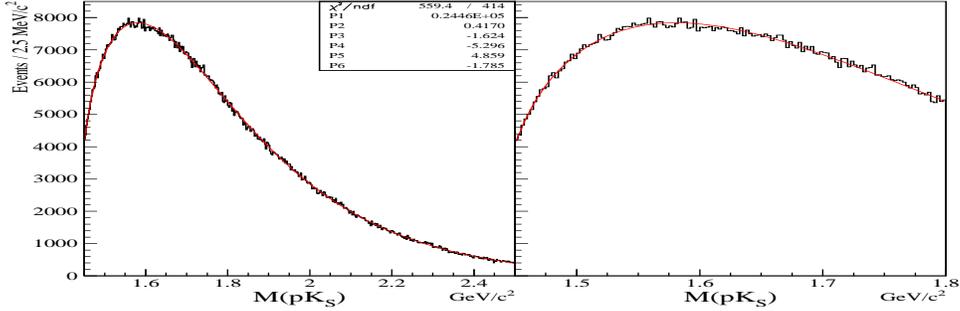}}
\caption{Fit to $M(pK_S^0)$ in search for $\Theta(1540)^+\!\rightarrow\!pK_S^0$.}
\label{fig:mpks}
\end{figure}

We search for $\Theta(1540)^+\!\rightarrow\!p K_S^0$ and measure the production 
relative to two similar decays, $K^*(892)^+\!\rightarrow\!K_S^0\pi^+$ and 
$\Sigma(1385)^\pm\!\rightarrow\!\Lambda^0\pi^\pm$.  The data is from 
events with a reconstructed $K_S^0\!\rightarrow\!\pi^+\pi^-$ or $\Lambda^0\!\rightarrow\!p\pi^-$\,\cite{focus_vee}. 
Selecting vee candidates within $2.5\,\sigma$ of the nominal mass we obtain
$63\times 10^6$ $K_S^0$ ($8 \times 10^6$ $\Lambda^0$)
candidates with 92\% (96\%) purity.  The remaining good 
quality tracks must form a good vertex (CL $>$ 1\%).
The proton candidate must pass stringent \v{C}erenkov ID cuts, reducing the misidentification rate to $\sim$0.
The $K^*(892)^-$ and $\Sigma(1385)^\pm$ are fit with a simple Breit-Wigner plus background of
$aq^b \exp{\left(cq + dq^2 + eq^3 +fq^4\right)}$ where $q$ is the energy release.  
We find $(8.29 \pm 0.01)\times 10^6$ $K^*(892)^-$, 
$(92 \pm 2)\times 10^3$ $\Sigma(1385)^+$, and $(146 \pm 3)\times 10^3$ $\Sigma(1385)^-$ signal events.
No pentaquark evidence is seen in the the $pK_S^0$ mass plot (Fig.\,\ref{fig:mpks}).  We obtain 95\% CL 
limits on the yield by determing how much the fitted yield must be increased to change the 
log-likelihood by 1.92 (continually minimizing the background parameters).  
This procedure is performed with a Breit-Wigner width of $0$ and
15\,MeV$\!/c^2$, both with Gaussian resolution varying from 2.36--3.07\,MeV$\!/c^2$.  
The maximum limit on the yield over the mass range 1.51--1.56\,GeV$\!/c^2$  is 
754 (2252) for $\Gamma$ of 0 (15) MeV$\!/c^2$.  To set cross section limits we generate pentaquarks the same as 
$\Sigma(1385)^+$.  Using \textsc{Pythia} production and FOCUS MC simulation of $K^*(892)^-$, $\Sigma(1385)^\pm$, 
and $\Theta(1540)^+$ events we convert the yield limits into cross section ratio limits.
For $1.51$$<$$M$$<$1.56\,GeV$\!/c^2$, the 95\% CL limit on the production of $\Theta(1540)^+$ relative to combined
$\Sigma(1385)^+$ and $\Sigma(1385)^-$ is 0.7\% (2.1\%) for 
$\Gamma$ of 0 (15) MeV$\!/c^2$.  Relative to $K^*(892)^-$, the limit is 0.06\% (0.17\%) for 
$\Gamma$ of 0 (15) MeV$\!/c^2$.  We account for all branching ratios and assume 
$B(\Theta(1540)^+\!\rightarrow\!pK_S^0) = 0.25$.

\section{Search for $\phi(1860)^{--}\!\rightarrow\! \Xi^- \pi^-$}

\begin{figure}
\centerline{\includegraphics[width=2.48in,height=1.8in]{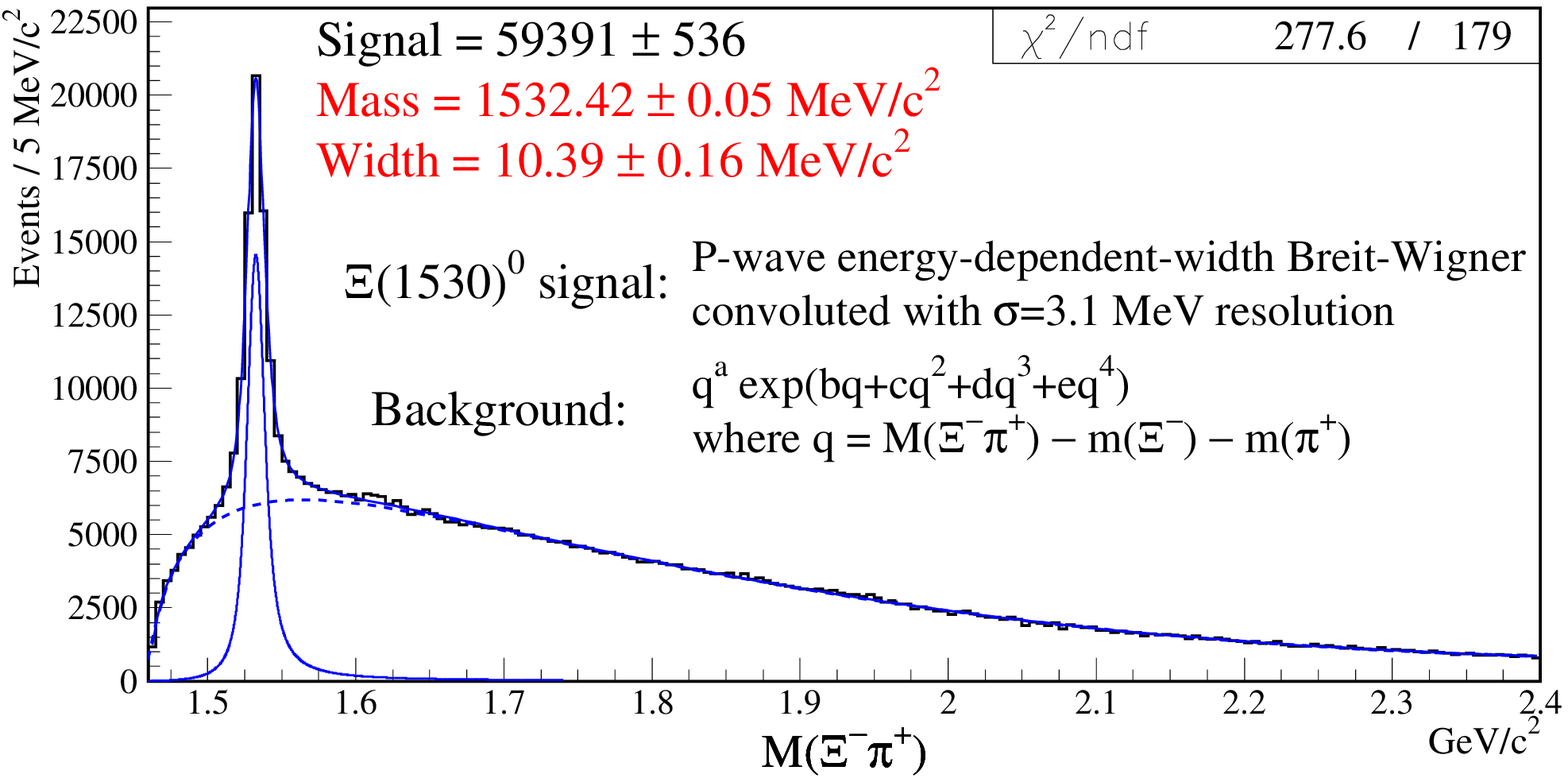}
\includegraphics[width=2.48in,height=1.8in]{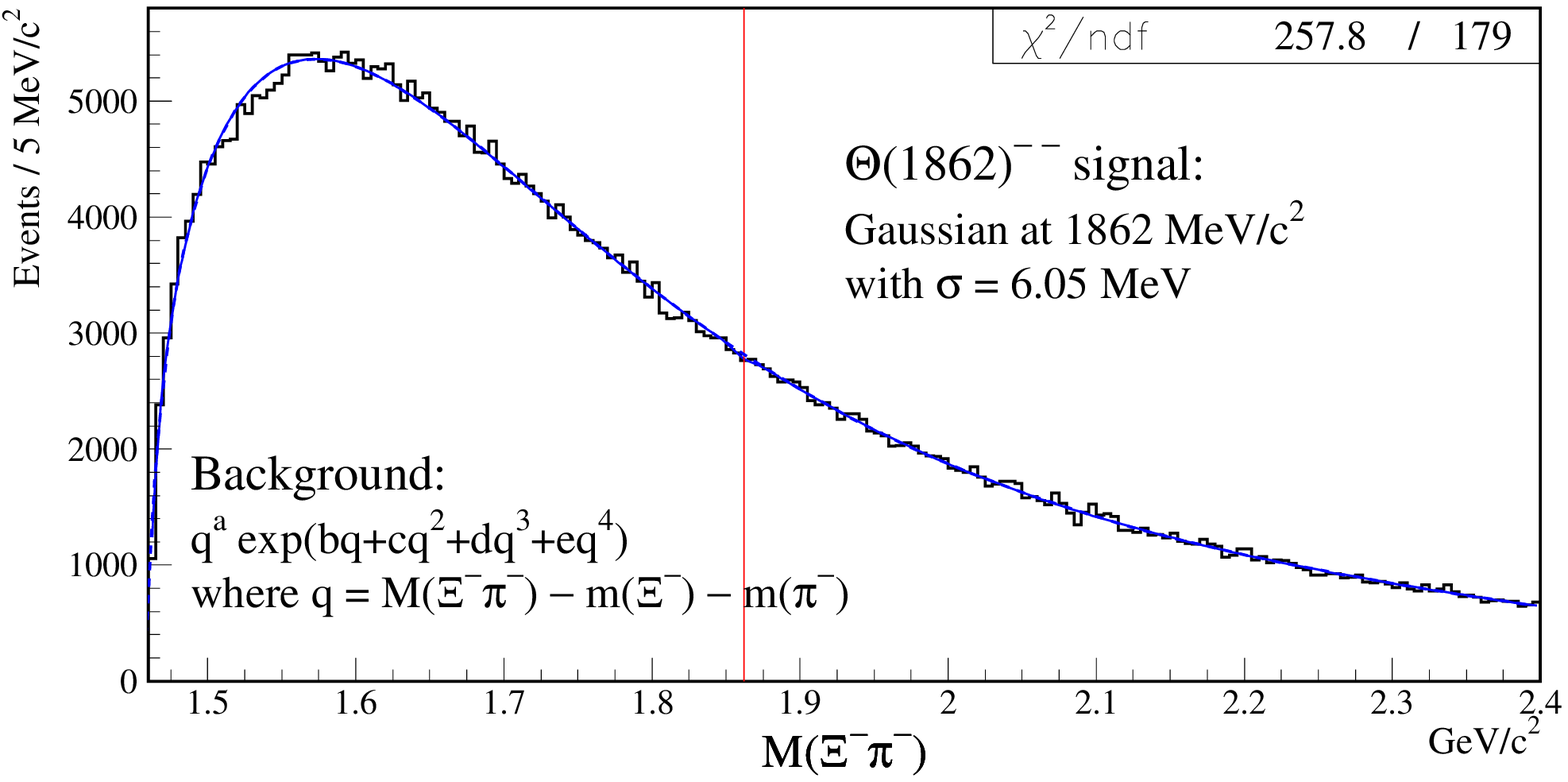}}
\caption{Fits to $M(\Xi^-\pi^+)$ (left) for $\Xi(1530)^0$ and $M(\Xi^-\pi^-)$ for $\phi(1860)^{--}$.}
\label{fig:xiplots}
\end{figure}

FOCUS reconstruction of $\Xi^- \!\rightarrow\! \Lambda^0\pi^-$ is described in 
Ref.\,\cite{focus_vee}.  We select 800,000 $\Xi^-$ candidates of which 75\% are signal.
We search for $\Xi(1530)^0\!\rightarrow\!\Xi^-\pi^+$ and the $S=-2$ pentaquark candidate
$\phi(1860)^{--}\!\rightarrow\!\Xi^-\pi^-$.  We require the production and $\Xi^-\pi^\pm$ vertices
have CL $\!>\!$ 1\% and separated by less than $2\sigma$.
The pion candidate must have a \v{C}erenkov signature consistent with a pion.
We find $59391 \pm 536$ $\Xi(1530)^0$ events
and no evidence for $\phi(1860)^{--}$ as shown in Fig.\,\ref{fig:xiplots}.  The yield 
upper limit calculated at a mass of $1.862$\,GeV$\!/c^2$ is 114 (170) for $\Gamma$ of 0 (15) MeV$\!/c^2$ with
resolution $\sigma = 6.05$\,MeV$\!/c^2$.  Assuming production like $\Xi(1530)^0$ we
find $\frac{\sigma(\phi(1860)^{--}) \times B(\phi(1860)^{--}\!\rightarrow\!\Xi^-\pi^-)}{\sigma(\Xi(1530)^0)} < 0.25\%\: (0.37\%)$ at
95\% CL for $\Gamma$ of 0 (15) MeV$\!/c^2$.

\section{Search for $\Theta_c(3099)^0\!\rightarrow\! D^{(*)-} p$}

\begin{figure}
\centerline{\includegraphics[width=2.48in,height=1.32in]{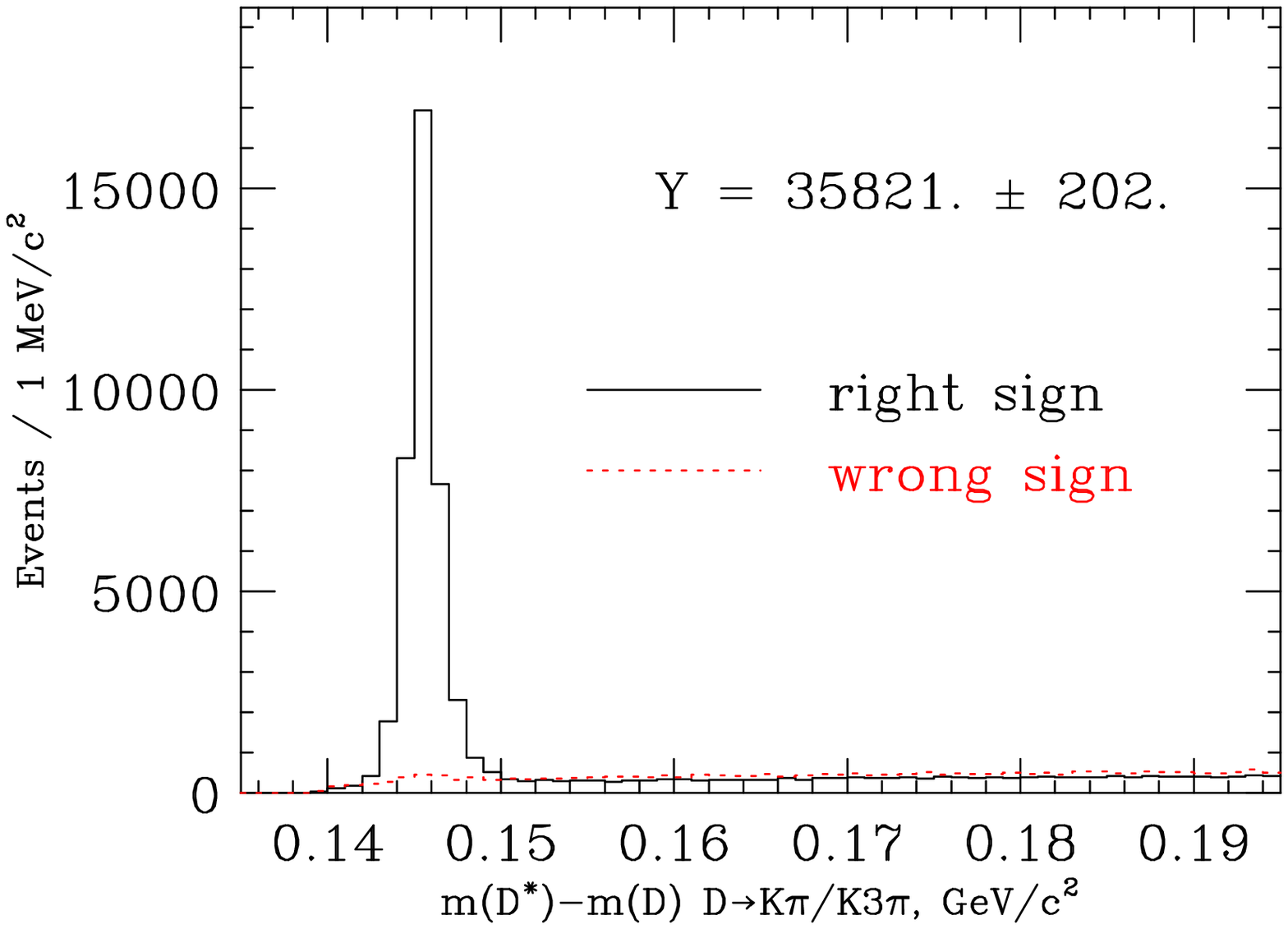}
\includegraphics[width=2.48in,height=1.32in]{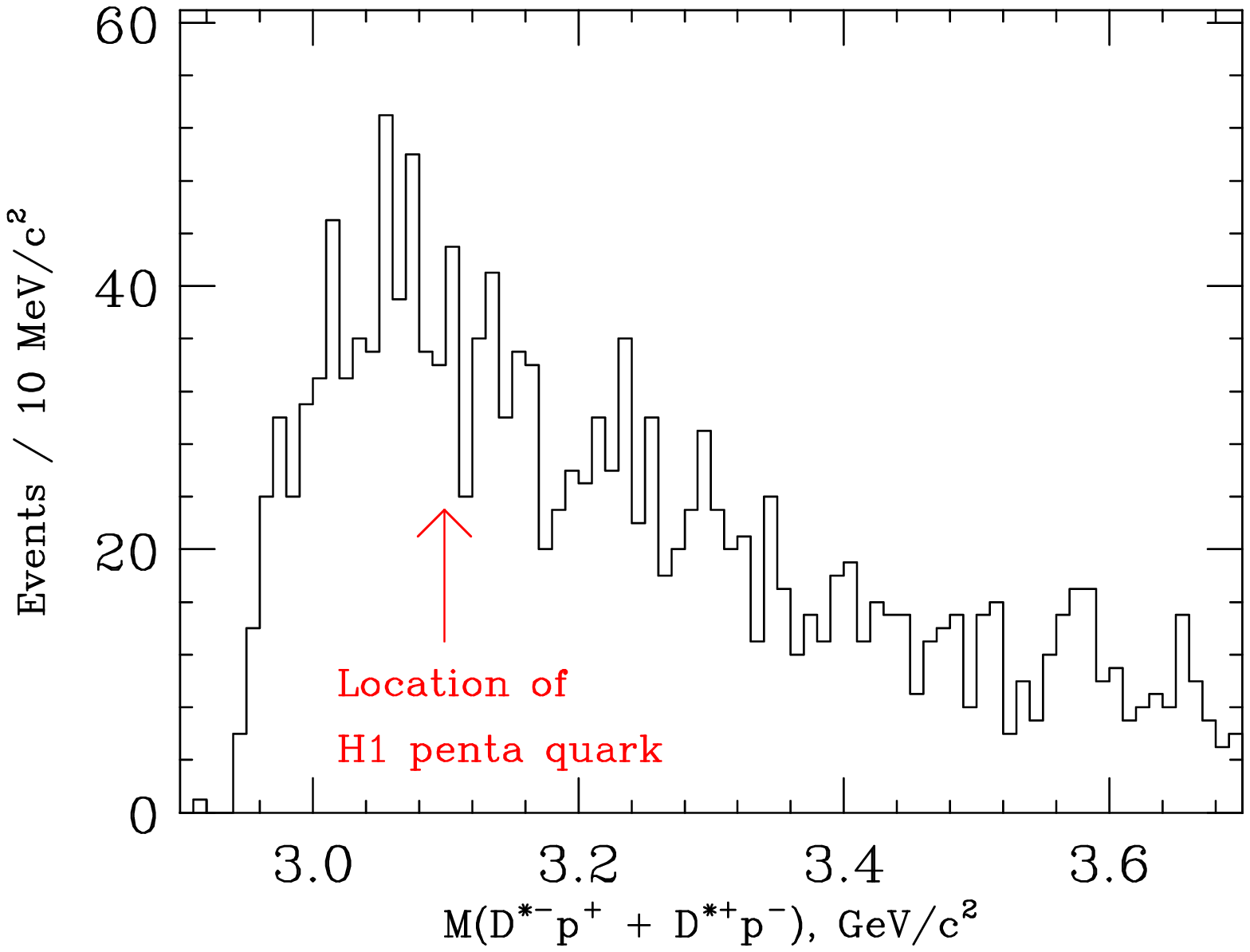}}
\centerline{\includegraphics[width=2.48in,height=1.32in]{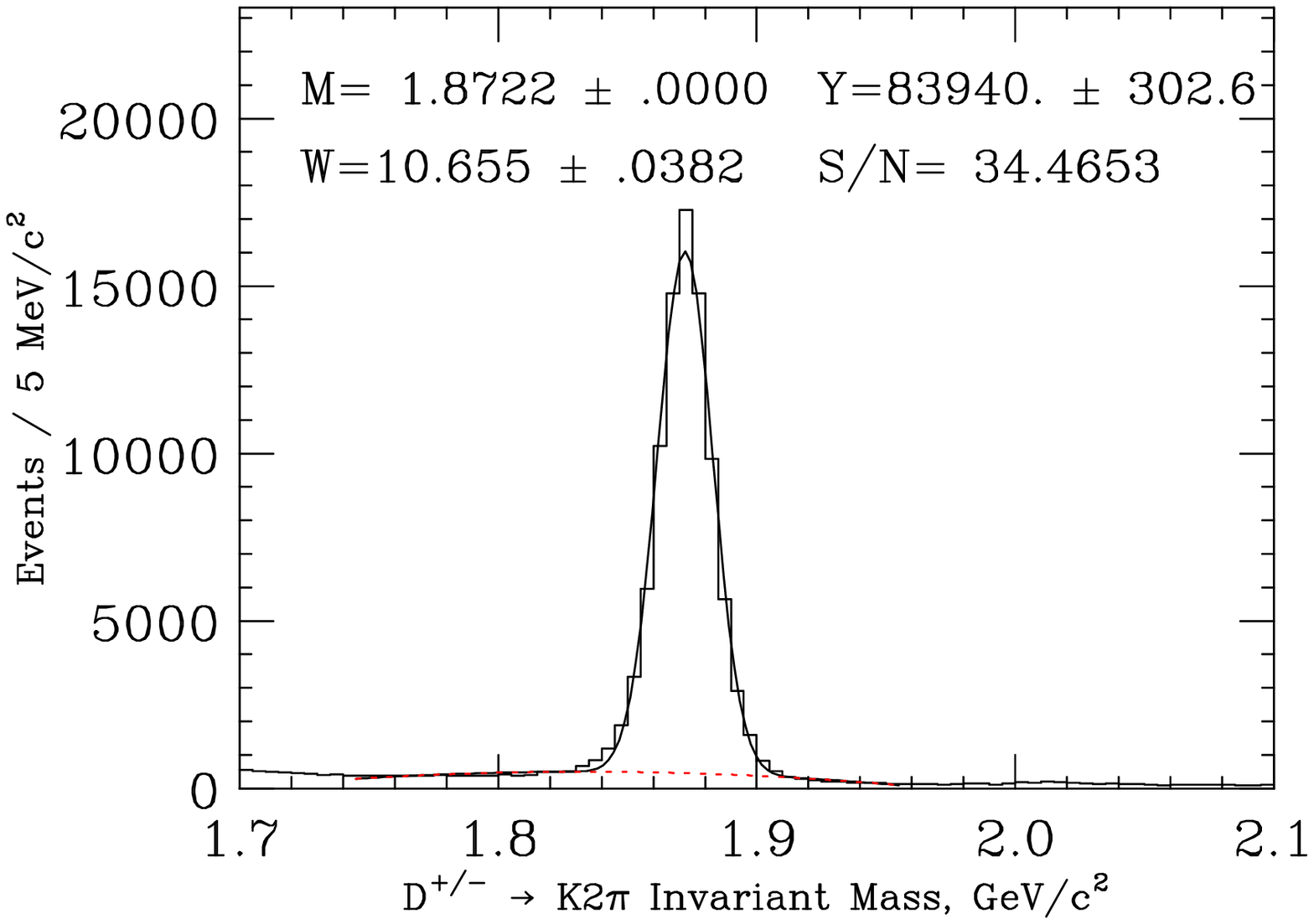}
\includegraphics[width=2.48in,height=1.32in]{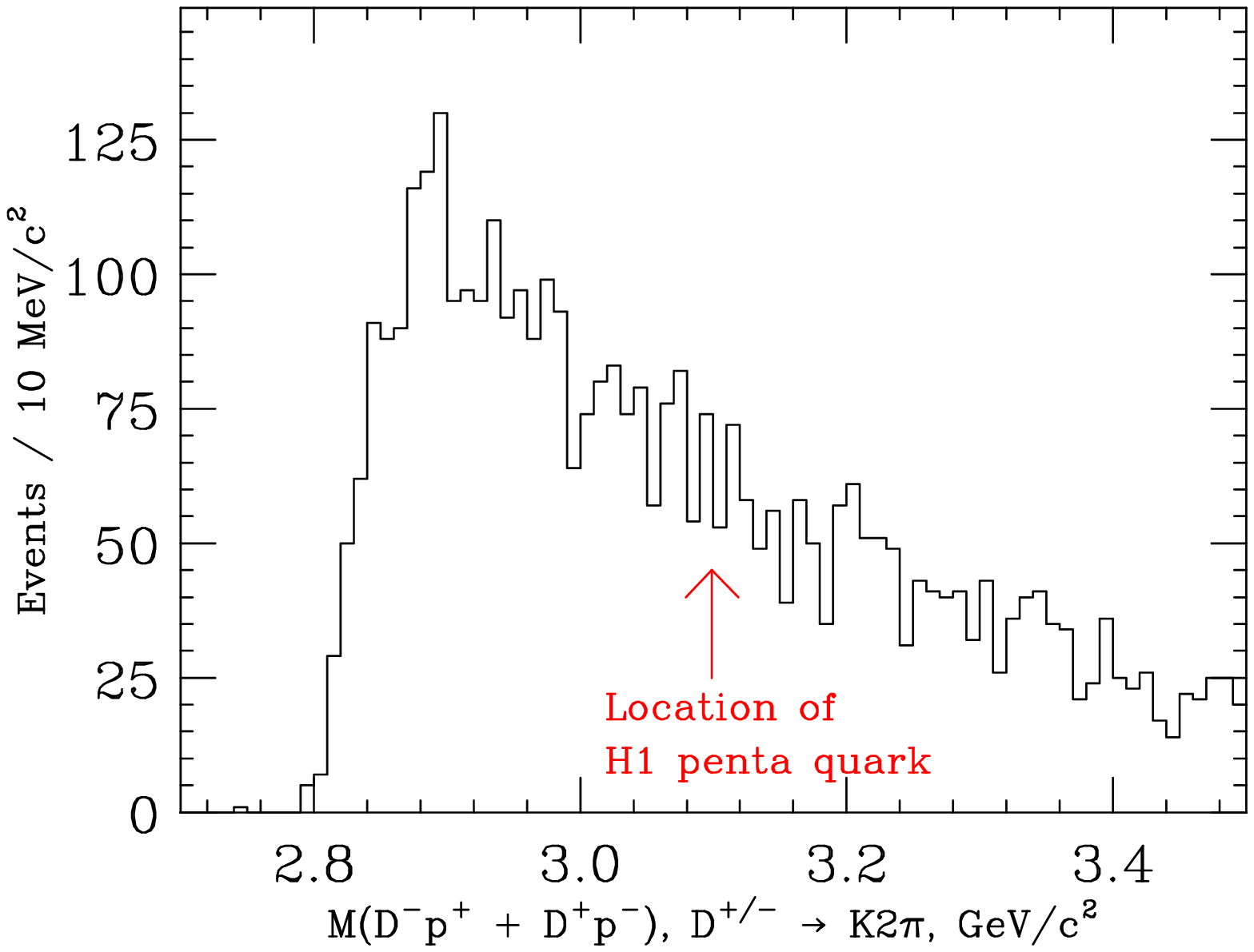}}
\caption{We add a proton to the $D^{*-}$ events and $D^-$ events (left) to search for $\Theta_c(3099)$ (right).}
\label{fig:thetacplots}
\end{figure}

Using standard FOCUS charm reconstruction techniques we obtain a clean
sample of $D^{*-}\!\rightarrow\!\overline{D}{}^0\pi^-$ 
$(\overline{D}{}^0\!\rightarrow\!K^+\pi^-)$ events and $D^-\!\rightarrow\!K^+\pi^-\pi^-$ events (Fig.~\ref{fig:thetacplots}).
Combining $35821 \pm 202$ $D^{*-}$ and $83940 \pm 303$ $D^-$ candidates with a
positively identified proton we find no evidence for a charm pentaquark as shown in Fig.~\ref{fig:thetacplots}.

\begin{thebibliography}{0}
\bibitem{leps} T.~Nakano {\it et al.} [LEPS Collaboration], \PRL(91,012002,2003).
\bibitem{diana} V.~V.~Barmin {\it et al.} [DIANA Collaboration], \PAN(66,1715,2003).
\bibitem{clas1} S.~Stepanyan {\it et al.}  [CLAS Collaboration], \PRL(91,252001,2003).
\bibitem{saphir} J.~Barth {\it et al.}  [SAPHIR Collaboration], \PL(B572,127,2003).
\bibitem{asratyan} A.~E.~Asratyan, {\it et al.}, \PAN(67,682,2004).
\bibitem{clas2} V.~Kubarovsky {\it et al.}  [CLAS Collaboration], \PRL(92,032001,2004).
\bibitem{hermes} A.~Airapetian {\it et al.}  [HERMES Collaboration], \PL(B585,213,2004).
\bibitem{svd} A.~Aleev {\it et al.}  [SVD Collaboration], arXiv:hep-ex/0401024
\bibitem{cosytof} M.~Abdel-Bary {\it et al.}  [COSY-TOF Collaboration], \PL(B595,127,2004).
\bibitem{zeus} S.~Chekanov {\it et al.}  [ZEUS Collaboration], \PL(B591,7,2004)
\bibitem{na49} C.~Alt {\it et al.}  [NA49 Collaboration],\PRL(92,042003,2004).
\bibitem{h1} A.~Aktas {\it et al.}  [H1 Collaboration], \PL(B588,17,2004).
\bibitem{focus_vee} J.~M.~Link {\it et al.}  [FOCUS Collaboration], \NIM(A484,174,2002).



\end{thebibliography}
\end{document}